\documentclass[aps,preprint]{revtex4}
\usepackage{amsfonts}
\usepackage{amssymb,epsfig}
\usepackage{latexsym}
\usepackage{amsmath}
\usepackage{graphicx}

\begin{document}

\title{Cosmic Behavior, Statefinder Diagnostic and $w-w^{\prime}$ Analysis
for\\ Interacting NADE model in Non-flat Universe}
\author{A. Khodam-Mohammadi\footnote{%
E-mail: \texttt{khodam@basu.ac.ir}}~~and\ M. Malekjani\footnote{%
E-mail: \texttt{malekjani@basu.ac.ir}}}
\address{Physics Department, Faculty of Science, Bu-Ali Sina
University, Hamedan, Iran}
\begin{abstract}
We give a brief review of interacting NADE model in non-flat
universe. we study the effect of spatial curvature $\Omega _{k}$,
interaction coefficient $\alpha $ and the main parameter of NADE,
$n$, On EoS parameter $w_{d}$ and deceleration parameter $q$. We
obtain a minimum value for $n$ in both early and present time, in
order to that our DE model crosses the phantom divide. Also in a
closed universe, changing the sign of $q$ is strongly dependent on
$\alpha$. It has been shown that the quantities $w_{d}$ and $q$ have
a different treatment for various spatial curvature. At last, we
calculate the statefinder diagnostic and $ w-w^{\prime}$ analysis in
non flat universe. In non flat universe, the statefinder trajectory
is discriminated by both $n$ and $\alpha$.
\end{abstract}

\maketitle

\section{Introduction\label{intro}}

The observational data of Ia supernova(SNIa) \cite{SN}, large scale
structure (LSS) \cite{LSS} and cosmic microwave background (CMB)
anisotropy \cite{CMB} show that the universe is undergoing an
accelerating expansion. It is believed that a dark energy with
negative pressure can drive this accelerated expansion. The dark
energy (DE) problem attracted a great deal
of attention in the last decade. Up to now, many models such as $\Lambda $%
CDM , the models with a scalar field and modified gravity had been proposed
\cite{Pad}.Also in the last decade the other models base on quantum field
theory such as holographic \cite{hol} and agegraphic (ADE) \cite{Cai1} dark
energy models are suggested. The latter is arisen from combining quantum
mechanics with general relativity directly. It is worthwhile to mention that
up to now, a completely successful quantum theory of gravity is not
available. There are two main problems in dealing with $\Lambda $CDM model
which are 'fine-tuning' and 'cosmic coincidence' problem \cite{copel}. The
ADE model, which is proposed by Cai \cite{Cai1}, is based on the line of
quantum fluctuations of spacetime, the so-called K\'{a}rolyh\'{a}zy relation
$\delta t=\lambda t_{p}^{2/3}t^{1/3}$, and the energy-time Heisenberg
uncertainty relation $E_{\delta t^{3}}\sim t^{-1}$. These relations enable
one to obtain an energy density of the metric quantum fluctuations of
Minkowski spacetime as follows \cite{Maz}
\begin{equation}
\rho _{q}\sim \frac{E_{\delta t^{3}}}{\delta t^{3}}\sim \frac{1}{%
t_{p}^{2}t^{2}}\sim \frac{m_{p}^{2}}{t^{2}}.  \label{ED}
\end{equation}%
Throughout this paper, we use the Plank unit ($\hbar =c=k_{B}=1$) ,
where $t_{p}=l_{p}=1/m_{p}$ are Plank's time, length and mass,
respectively. In ADE, this energy density is considered as density
of dark energy component, $\rho _{d}$, of spacetime. By considering
a Friedmann-Robertson-Walker (FRW) universe, due to effect of
curvature, one should introduce a numerical factor $3n^{2}$ in
(\ref{ED}) \cite{cai2, cai3}.

By making a model that considering DE, independent of the other
mater fields, one can study the evolution of characteristics of dark
energy of spacetime very well. The recent observational data from
the Abell Cluster $A586$ supports the interaction between dark
matter and dark energy \cite{bertol}. However, the strength of this
interaction is not exactly identified \cite{feng}. Also, nowadays,
many authors are interested to consider non-flat FRW universe
\cite{shikh,karami}. The tendency of a closed universe is shown in a
suite CMB experiments \cite{Sie}. Besides of it, the measurements of
the cubic correction to the luminosity-distance of supernova
measurements reveal a closed universe \cite{Caldwell}. In accordance
of all mentioned above, we prefer to consider a model including dark
matter and dark energy for a non-flat FRW universe.

The interacting new ADE (NADE) which is a new version of ADE with a
conformal time instead of a cosmic time in FRW metric. The
motivation of this new model is for that the original ADE model
cannot explain the matter-dominated era \cite{cai4}. Recently, the
interacting NADE model has been investigated and some cosmological
quantities such as deceleration parameter $q$,
evolution behavior of fractional dark energy density $\Omega _{d}^{\prime }$%
, and equation of state (EOS) parameter $w_{d}=P_{d}/\rho _{d}$, are
obtained \cite{wei,zhang}.

The next step beyond Hubble parameter $H=\dot{a}/a$ and
$q=-\ddot{a}/aH^{2}$, is to consider a new quantity contains
$\dddot{a}$. A pair quantities which have been introduced by Sahni
et al. and Alam et al. \cite{Sahni}, are
called statefinder pair \{r,s\}, as%
\begin{equation}
r=\frac{\dddot{a}}{aH^{3}},~~~~~~~s=\frac{r-1}{3(q-1/2)},  \label{rspair}
\end{equation}%
The statefinder pair is a geometrical diagnostic tool which is constructed
directly from a spacetime metric. The importance of such pair is to
distinguish of the cosmological evolution behaviors of dark energy models
with the same values of $H_{0}$ and $q_{0}$ at the present time. At future
by combining the data of Supernova acceleration probe (SNAP) with
statefinder diagnosis, we may choose the best model of dark energy. Up to
now, many authors have investigated statefinder trajectories for standard $%
\Lambda $CDM model and quintessence \cite{r9,r10}, interacting quintessence
models \cite{r12,r13}, chaplygin gas, the holographic dark energy models
\cite{r14,r15}, the holographic dark energy model in non-flat universe \cite%
{r16}, the phantom model \cite{r18}, the tachyon \cite{r22}, the ADE model
with and without interaction \cite{wei} and the interacting NADE model in
flat universe \cite{zhang}. They had shown the statefinder diagnosis is a
useful tool for discrimination between various dark energy models. In
addition to the statefinder geometrical diagnostic, the another tool to
distinguish between the different models of dark energy is $w-w^{\prime }$
analysis which is used extensively in the literature \cite{wei,wwp}.

In this paper, in addition of cosmic behavior investigation, we
study the statefinder trajectories and $w-w^{\prime }$ analysis for
interacting NADE model in a non-flat FRW universe.

\section{COSMIC EVOLUTION IN NON-FLAT UNIVERSE WITH INTERACTING NADE\label%
{theory}}

As we mentioned in Sec. (\ref{intro}), the energy density can be defined in
ADE model as
\begin{equation}
\rho _{d}=\frac{3n^{2}m_{p}^{2}}{T^{2}}  \label{adedens}
\end{equation}%
where the cosmic time $T$ is defined as the age of the universe%
\begin{equation}
T=\int dt=\int_{0}^{a}\frac{da}{Ha}.
\end{equation}%
Introducing a conformal time $\eta $ which is defined as $dt=ad\eta $, the
FRW universe is modified as
\begin{equation}
ds^{2}=dt^{2}-a^{2}dx^{2}=a^{2}(d\eta ^{2}-dx^{2}).
\end{equation}%
Therefore, by substituting the time scale $\eta $ in Eq.(\ref{adedens}), one
can obtain the energy density of NADE model as%
\begin{equation}
\rho _{d}=\frac{3n^{2}m_{p}^{2}}{\eta ^{2}};\qquad \eta =\int \frac{dt}{a}%
=\int_{0}^{a}\frac{da}{Ha^{2}};\qquad \dot{\eta}=\frac{1}{a}.  \label{endend}
\end{equation}%
The corresponding fractional energy density is%
\begin{equation}
\Omega _{d}=\frac{n^{2}}{H^{2}\eta ^{2}}.  \label{Fendend}
\end{equation}%
The Friedmann equation of a non-flat FRW universe containing a new
agegraphic
dark energy and pressureless matter (baryons and dark matter) is%
\begin{equation}
H^{2}+\frac{k}{a^{2}}=\frac{1}{3m_{p}^{2}}(\rho _{m}+\rho _{d})  \label{Freq}
\end{equation}%
where $k=1,0,-1$ is curvature parameter corresponding to closed, flat and
open universe, respectively. Some recent observations reveal a closed
universe with a present small fractional energy density $\Omega
_{k0}=1/H_{0}^{2}\simeq 0.02$ \cite{Bennet}. Also we can write the Friedmann
equation (\ref{Freq}) in to another form with respect to fractional energy
density $\Omega _{i}=\rho _{i}/\rho _{c}$, with $\rho _{c}=3m_{p}^{2}H^{2}$.
Then we have
\begin{equation}
\Omega _{m}+\Omega _{d}=1+\Omega _{k}.  \label{Freq2}
\end{equation}%
The continuity equations including an interaction term between dark matter
and dark energy become%
\begin{eqnarray}
\dot{\rho _{m}}+3H\rho _{m} &=&Q, \\
\dot{\rho _{d}}+3H(\rho _{d}+p_{d}) &=&-Q,  \label{contd}
\end{eqnarray}%
where $p_{d}$ is dark energy pressure which is given by equation of
state (EoS), $w_{d}=p_{d}/\rho _{d}$. Three forms of $Q$ which have
been
extensively used in literatures \cite{cai4, zhang, shikh} are%
\begin{equation}
Q=9\alpha _{i}m_{p}^{2}H^{3}\Omega _{i};\qquad \Omega _{i}=\left\{
\begin{array}{ll}
\Omega _{d}; & i=1 \\
\vspace{0.75mm}\Omega _{m}; & i=2 \\
\Omega _{d}+\Omega _{m}; & i=3%
\end{array}%
\right. .  \label{qform}
\end{equation}%
Differentiating Eq. (\ref{Fendend}) and using Eqs. (\ref{adedens}), (\ref%
{Freq}) and (\ref{Freq2}), the derivative of $\Omega _{d}$ can be calculated
as%
\begin{eqnarray}
\Omega _{d}^{\prime } &=&\frac{\dot{\Omega}}{H}=-2\Omega _{d}\left[ \frac{%
\dot{H}}{H^{2}}+\frac{\sqrt{\Omega _{d}}}{na}\right] ;  \label{OmP} \\
\frac{\dot{H}}{H^{2}} &=&-\frac{\Omega _{d}^{3/2}}{na}-\frac{3}{2}(1-\Omega
_{d})-\frac{\Omega _{k}}{2}+\frac{Q_{c}}{2},  \label{HD}
\end{eqnarray}%
where prime denotes the derivative with respect to $\ln a$ and $%
Q_{c}=Q/H\rho _{c}=3\alpha _{i}\Omega _{i}.$ The relations (\ref{OmP}) and (%
\ref{HD}), also has been obtained in \cite{shikh} for third interaction form
of $Q.$

Using the relation (\ref{HD}) in (\ref{OmP}), we obtain a normal
differential equation for $\Omega _{d}$ as%
\begin{equation}
\Omega _{d}^{\prime }=\Omega _{d}\left[ (1-\Omega _{d})(3-2\frac{\sqrt{%
\Omega _{d}}}{na})+\Omega _{k}-Q_{c}\right] ,  \label{Oddiff}
\end{equation}%
where $\Omega _{k}$ is given by
\begin{equation}
\Omega _{k}=\frac{a\gamma (1-\Omega _{d})}{1-a\gamma }.  \label{Ok}
\end{equation}%
Hear $\gamma $ is satisfied in the following equation%
\begin{equation}
\frac{\Omega _{k}}{\Omega _{m}}=a\frac{\Omega _{k0}}{\Omega _{m0}}=a\gamma .
\end{equation}%
From Eqs. (\ref{endend}), (\ref{Fendend}) and (\ref{contd}), the EoS
parameter can be obtained as%
\begin{equation}
w_{d}=-1-Q_{cd}+\frac{2\sqrt{\Omega _{d}}}{3na},  \label{EoS}
\end{equation}%
where $Q_{cd}=Q_{c}/(3\Omega _{d})$. The evolution behavior of $w_{d}$ (\ref%
{EoS}) by using (\ref{Oddiff}), is given by
\begin{equation}
w_{d}^{\prime }=+\frac{\sqrt{\Omega _{d}}}{3na}\left[ 1+\Omega _{k}-3\Omega
_{d}-2\frac{\sqrt{\Omega _{d}}}{na}(1-\Omega _{d})-Q_{c}\right]
-Q_{cd}^{\prime }.  \label{wp}
\end{equation}%
Also the total EoS parameter is obtained as%
\begin{equation}
w_{tot}=\frac{p_{tot}}{\rho _{tot}}=\frac{\Omega _{d}w_{d}}{1+\Omega _{k}}.
\end{equation}%
To achieve an accelerated expansion, it is required that $w_{tot}$
$<-1/3$.
Therefore at present time in a closed universe ($a=1,\Omega _{k0}=0.02,$ $%
\Omega _{d0}=0.72$), the minimum value of $n$ with $\alpha _{i}=0.1$ for
various forms of $Q,$ can be obtained as%
\begin{equation}
n\geq \left\{
\begin{array}{ll}
0.901; & i=1 \\
\vspace{0.75mm}0.993; & i=2 \\
0.845; & i=3%
\end{array}%
\right.
\end{equation}

In interacting NADE, based on Eq. (\ref{EoS}) , we see that at
present time, $w_{d}$ can cross the phantom divide ($w<-1$) provided
\begin{equation}
n\alpha \geq \left\{
\begin{array}{ll}
0.566; & i=1 \\
1.358; & i=2 \\
0.399; & i=3%
\end{array}%
\right.   \label{nalpha}
\end{equation}%
The third value of (\ref{nalpha}), has been also expressed in
\cite{shikh}. The relation (\ref{nalpha}) show that for the same
value of $\alpha $, $n_{i} $ is increasing by moving from $i=3$ in
to $i=1$ and then $i=2.$

The deceleration parameter $q$ can be calculated from Eqs. (\ref{OmP}) and (%
\ref{Oddiff}) as%
\begin{eqnarray}
q &=&-\frac{\dot{H}}{H^{2}}-1=\frac{(1+\Omega _{k})}{2}+\frac{3}{2}\Omega
_{d}w_{d}  \label{q} \\
&=&\frac{\Omega _{d}^{3/2}}{na}-\Omega _{d}+\frac{1-\Omega _{d}}{2(1-a\gamma
)}-\frac{Q_{c}}{2}  \label{q2}
\end{eqnarray}%
$\allowbreak $

\section{STATEFINDER DIAGNOSTIC OF NADE IN INTERACTING NON-FLAT UNIVERSE
\label{STF}}

Now we find the statefinder pair \{r,s\}, which was expressed in Sec.\ref%
{intro}. From the definition of $q$ and $H$, the parameter $r$ (\ref{rspair}%
) can be written as
\begin{equation}
r=\frac{\ddot{H}}{H^{3}}-3q-2.  \label{r}
\end{equation}%
Using Eqs. (\ref{HD}), (\ref{EoS}) and (\ref{q}) we have%
\begin{equation}
\frac{\ddot{H}}{H^{3}}=\frac{9}{2}+\frac{9}{2}\Omega
_{d}w_{d}(w_{d}+Q_{cd}+2)-\frac{3}{2}\Omega _{d}w_{d}^{\prime }+\frac{5}{2}%
\Omega _{k}.  \label{Hdd}
\end{equation}%
Hence, the Eq. (\ref{r}) can be obtained as%
\begin{equation}
r=1+\Omega _{k}+\frac{9}{2}\Omega _{d}w_{d}(w_{d}+Q_{cd}+1)-\frac{3}{2}%
\Omega _{d}w_{d}^{\prime }.  \label{r2}
\end{equation}
In a non flat universe, Evans et al. \cite{Evans} generalize the
definition of parameter $s$ (\ref{rspair}) as
\begin{equation}
s=\frac{r-\Omega _{tot}}{\frac{3}{2}(q-\frac{\Omega _{tot}}{2})},  \label{sg}
\end{equation}%
where the total fractional energy density is $\Omega _{tot}=\Omega
_{m}+\Omega
_{d}=1+\Omega _{k}$.  Therefore from this new definition we have%
\begin{equation}
s=1+w_{d}+Q_{cd}-\frac{w_{d}^{\prime }}{3w_{d}}.  \label{sk}
\end{equation}
The relations (\ref{Hdd}),(\ref{r2}) and (\ref{sk}) reduced to (35),
(37) and (38) of Ref. \cite{zhang} in the limiting case of flat
universe. By omitting $w_{d}^{\prime }$ between (\ref{r2}) and
(\ref{sk}), we find $r$ in terms of $s $ as follows
\begin{equation}
r=1+\Omega _{k}+\frac{9}{2}s\Omega _{d}w_{d}.  \label{r3}
\end{equation}

\section{Numerical results\label{NR}}

In this section, first we give the complete numerical description of the
NADE model and then examine the NADE model with statefinder diagnostic tool
and $w-w^{\prime }$ analysis. Here we consider the first case of interaction
form, $Q=Q_{1}=9\alpha _{1}m^{2}H^{3}\Omega _{d}$, in (\ref{qform}) with $%
\alpha =\alpha _{1}$. In this case, the differential equation for
$\Omega _{d} $ (\ref{Oddiff}) can be reduced as
\begin{equation}
\Omega _{d}^{\prime }=\Omega _{d}[(1-\Omega _{d})(3-\frac{2\sqrt{\Omega _{d}}%
}{na})-3\Omega _{d}\alpha +\Omega _{k}]  \label{Omeg_d}
\end{equation}
Substituting $Q_{1}$ in Eqs.(\ref{EoS}), (\ref{wp}), (\ref{r2}) and (\ref{sk}%
), yields the following equations:
\begin{equation}
w_{d}=-1+\frac{2\sqrt{\Omega _{d}}}{3na}-\alpha   \label{w_d}
\end{equation}
\begin{equation}
w_{d}^{\prime }=\frac{\sqrt{\Omega _{d}}}{3na}[(1+\Omega _{k})-3\Omega
_{d}(1+\alpha )-\frac{2\sqrt{\Omega _{d}}}{na}(1-\Omega _{d})]  \label{wp_d}
\end{equation}
\begin{equation}
r=1+\Omega _{k}+\frac{9}{2}\Omega _{d}w_{d}(w_{d}+\alpha +1)-\frac{3}{2}%
\Omega _{d}w_{d}^{\prime }.  \label{state_r}
\end{equation}
\begin{equation}
s=1+w_{d}+\alpha -\frac{w_{d}^{\prime }}{3w_{d}}.  \label{state_s}
\end{equation}
Because of vanishing $\Omega_{d}$ at $a\rightarrow 0$, the second
term of Eq.(\ref{w_d}) can be larger or smaller than unity at this
time. This term is examined by numerical analysis and we see that it
is much lower than one for higher values of model parameter $n$.
Hence, NADE model can cross the phantom divide ($w_{d}<-1$) at
$a\rightarrow 0$, independent of the contribution of the spatial
curvature of the universe.

On the other hand, at the late time (e.g., $a\rightarrow \infty $)
when $\Omega _{d}\rightarrow 1$, $w_{d}$ tends to $-1$ and NADE
model mimics the cosmological constant. Here we focus on
Eq.(\ref{w_d}) in more details. $w_{d}$ crosses the phantom divide
($w_{d}<-1$) when $2\sqrt{\Omega _{d}} /3na<\alpha $, otherwise
$w_{d}$ can not cross the phantom divide. In Sect.\ref{theory}, we
obtained a condition for $n\alpha >0.566$ at present time for
crossing the phantom divide. For the best value of $n=2.7$ which has
been obtained from astronomical data for NADE model, we should set
$\alpha>0.2$ to have $w_{d}<-1$ at present time. However, the
situation is changed at the early time, since both $a$ and $\Omega
_{d}$ tend to zero at that time. For example the values $n=4$ and
$\alpha =0.1$ satisfies the relation $ 2\sqrt{\Omega
_{d}}/3na<\alpha $ at the early times and $w_{d}$ crosses the
phantom divide. while the condition $2\sqrt{\Omega _{d} }/3na<\alpha
$ can not be satisfied for the values $n=3$,$\alpha =0.1$ and the
phantom divide can not be achieved at the early times in this case.

Fig.(1) shows the evolution of $w_{d}$ of interacting NADE model in
terms of scale factor. In Fig.(1-a) we illustrate the evolution of
$w_{d}$ for fixed model parameters $n=4$ and $\alpha =0.1$, in open,
flat and closed universe. All three cases give the phantom divide at
the early time and mimic the cosmological constant at the late time.
In Figs.(1-b,1-c), the dependence of the evolution of $w_{d}$ on the
model parameters $n$ and $\alpha $ are investigated. Here we choose
the closed universe with the present spatial curvature $\Omega
_{k0}=0.02$. In Fig.(1-b), by fixing $\alpha =0.1$, we vary the
parameter $n$ as 3, 4 and 5. In the case of $n=3$ the interacting
NADE model can not cross the phantom divide, while for $n=4$ and
$n=5$ the phantom divide can be achieved. In Fig.(1-c) we fix $n=4$
and vary $\alpha $ as $\alpha =0.0$ , $ \alpha =0.1$ and eventually
$\alpha =0.15$. It can be seen that the phantom divide is archived
for $\alpha =0.1$ and $\alpha =0.15$ and it can not be access for
$\alpha =0.0$ (the NADE model without interaction).

The other cosmological parameter which we demonstrate, is the
deceleration parameter $q$. The parameter $q$ in NADE model for
non-flat universe is given by Eq.(\ref{q}). In the early time, where
$\Omega _{d}\rightarrow 0$ and $\Omega _{k}\rightarrow 0$, the
parameter $q$ converges to $1/2$, whereas the universe has been
dominated by dark matter. In Fig(2), we show the evolution of $q$ as
a function of cosmic scale factor for different model parameters of
NADE model and also for various contribution of spatial curvature of
the universe. In Fig.(2-a), the dependence of the evolution of $ q$
on the spatial curvature of the universe ($k=0,1,-1$) is sketched
for $n=4 $ and $\alpha =0.1$. In this model, the deceleration
parameter $q$ crosses the boundary $q=0$ from $q>0$ to $q<0$. This
implies that the universe undergoes decelerated expansion at the
early time and later starts accelerated expansion. The transition
from decelerated expansion to the accelerated expansion occurs
gradually from closed, flat and open universe. However, the
difference between them is very little, but we can interpret that
the transition occurs earlier in closed universe. In Figs.(2-b,2-c)
the evolution of $q$ in terms of the scale factor is plotted for
different values of $n$ and $\alpha $ in the case of closed universe
with $\Omega _{k0}=0.02$.  In Fig.(2-b), we set $\alpha =0.1$ and
change $n$ as 3, 4 and 5. Here the change on the sign of $q$ is
taken place at similar $a$ for all values of $n$. It should be noted
that in the accelerated universe ($q<0$), the parameter $q$ is
smaller for higher values of $n$ while in the decelerated universe
($q>0$), $q$ is larger for higher values of $n$. In Fig.(2-c), by
fixing $n=4$, and changing $ \alpha $ as $0.0$, $0.1$ and $0.15$ the
behavior of $q$ is studied. Here the universe starts accelerated
expansion earlier when $\alpha $ is more.

At following, we calculate the evolution trajectories in the
statefinder planes and analyze the interacting NADE model in
non-flat universe with statefinder point of view. The standard
$\Lambda $CDM model in non flat universe corresponds to the fixed
point ($s=0$,$r=1+\Omega _{k}$) in the $s-r$ plane \cite{Evans}. One
way to test the ability of a given dark energy model is the
deviation value of the model from the fixed point ($s=0$,$r=1+\Omega
_{k} $) in $s-r$ diagram. Let us start with
Eqs.(\ref{wp_d},\ref{state_s} and \ref {state_r})which describe the
evolution of statefinder parameters $r$, $s$ and also $w_{d}^{\prime
}$. It is easy to see that in the early time, $ w_{d}^{\prime
}\rightarrow 0,$ and from ( \ref{state_r}) and(\ref{state_s}),
$s\rightarrow 0$ and $r\rightarrow 1$. Also it is worth to estimate
the values of $\omega _{d}^{\prime }$, $s$ and $r$ at late time when
$a\rightarrow \infty $. From Eqs.( \ref{state_r})
and(\ref{state_s}), we obtain $s\sim \alpha $ and $r\sim 1$ at late
time. So the statefinder parameters ($s$,$r$) reach to the ($\alpha
$,$1$) at the late time.

In Fig.(3) the evolution trajectories of statefinder of interacting NADE
model with interaction form of $Q_{1}$ is plotted. In Fig.(3-a) , the
evolution trajectories is plotted for different closed, flat and open
universe for fixed parameters $n=4$ and $\alpha =0.1$. The present values of
the statefinder parameters $s_{0}$ and $r_{0}$ is denoted by circle ($%
s_{0}=0.154$,$r_{0}=0.535$), star ($s_{0}=0.154$,$r_{0}=0.515$) and square ($%
s_{0}=0.153$,$r_{0}=0.497$) symbols for closed, flat and open
universe, respectively. It should be noted that the evolution
trajectories start form fixed point ($s=0,r=1$) at the early time,
as mentioned above. It can be seen that the different curvatures
will lead to different evolutionary behavior in the statefinder
plane, starting from the same fixed point ($s=0$,$r=1$) at the early
time. The curvature will affect the today's value of statefinder
parameter. Fig.(3-a) shows that the distance to $\Lambda $CDM fixed
point in closed universe is shorter than of obtained in flat
universe and both of them is shorter than that distance in open
universe. Also in closed universe the value of $r$ is the largest,
while $s$ is the smallest at present time.

Figs.(3-b \& 3-c) indicate the dependence of the evolution
trajectories of statefinder diagnostic on $n$ and $\alpha $ in
closed universe. Fig.(3-b) shows the influence of the variation of
$n$ on the evolution trajectories of statefinder for fixed parameter
$\alpha =0.1$. Here the evolution trajectories is calculated for
$n=3,4$ and $5$. Symbols on the curves represent the present value
of statefinder. The circle indicates ($ s_{0}=0.205$,$r_{0}=0.406$)
for $n=3$, the star symbol indicates ($ s_{0}=0.154$,$r_{0}=0.535$)
in the case $n=4$ and the square denotes ($
s_{0}=0.123$,$r_{0}=0.620$) for $n=5$. Increasing the parameter $n$
will lead to shorter distance between present values ($s_{0},r_{0}$)
and fixed $ \Lambda $CDM in this diagram. Also we can see that the
higher value of $n$ makes the larger value of $r$ and smaller value
of $s$.\newline In Fig.(3-c), we redo the previous calculation in
Fig.(3-b) for fixed parameter $n=4$ and varying parameter $\alpha $
as $0.0$, $0.1$, $0.15$. The variation of $\alpha $ also change the
evolution trajectories of statefinder. The present values of
statefinder for different values of $ \alpha $ are:
($s_{0}=0.162$,$r_{0}=0.564$) for $\alpha =0.0$, ($s_{0}=0.154$
,$r_{0}=0.535$) for $\alpha =0.1$ and ($s_{0}=0.151$,$r_{0}=0.521$)
for $ \alpha =0.15$. We can see that the higher value of $\alpha $
makes the smaller value of $r$ and also the smaller value of $s$. It
is worth noting that in the case of flat universe, only the
parameter $\alpha$ can discriminate the evolution trajectory in
$s-r$ plane and the parameter $n$ can only sperate toady's value of
$s$ and $r$(see Figs.(3 \& 4) of \cite{zhang}). Here in non flat
case, both parameters $n$ and $\alpha$ can discriminate the
evolution trajectories in $s-r$ plane (see Figs.3-b \& 3-c).\\
At last, we study the interacting NADE model in non flat universe
using the $w-w^{\prime }$ analysis. In this analysis, the standard
$\Lambda $CDM model corresponds to the fixed point
($w_{d}=-1$,$w_{d}^{\prime }=0$) in the $w-w^{\prime }$ plane. The
evolution of $w_{d}$ and $w_{d}^{\prime }$ is given by
Eqs.(\ref{w_d}, \ref{wp_d}). In Fig.(4) the evolution trajectories
of $ w_{d}^{\prime }$ in $w-w^{\prime }$ plane is shown for
different parameters and various curvatures. In Fig.(4-a), we show
the evolution trajectories of $ w_{d}^{\prime }$ , by fixing the
parameters $n=4$ and $\alpha =0.1$ for different spatial curvatures.
Here we see that the various spatial curvatures gives the different
evolutionary behavior in the $w-w^{\prime }$ plane. The evolution
trajectory of different curvatures converges to the fixed point
($w_{d}=-1.1,w_{d}^{\prime }=0.0$) at late time. The present value
of $w_{d}$ and $w_{d}^{\prime }$ are: ($w_{d}^{0}=-0.957$,$
w_{d}^{\prime 0}=-0.104$) in closed,
($w_{d}^{0}=-0.957$,$w_{d}^{\prime 0}=-0.106$) in flat and
($w_{d}^{0}=-0.957$,$w_{d}^{\prime 0}=-0.107$) in open
universe.\newline In Figs.(4-b, 4-c), the evolution trajectories in
$w-w^{\prime }$ plane is discussed for the case of closed universe.
In Fig.(4-b), by fixing $\alpha =0.1$, we vary the parameter $n$ as
$3,4$ and $5$. The today's value of $ w_{d}^{0},w_{d}^{\prime 0}$ is
denoted by symbols on the lines. The circle symbol shows the present
value ($w_{d}^{0}=-0.910,w_{d}^{\prime 0}=-0.142$) in the case of
$n=3$, star ($w_{d}^{0}=-0.957,w_{d}^{\prime 0}=-0.104$) for $ n=4$
and square ($w_{d}^{0}=-0.986,w_{d}^{\prime 0}=0-0.082$) for $n=5$.
Increasing the parameter $n$ will lead to closing the present values
of $ w_{d}^{0},w_{d}^{\prime 0}$ tend to the fixed point ($-1,0$) in
$w-w^{\prime }$ plane. The higher value of $n$ obtains the smaller
values of $w_{d}$ and the bigger value of $w^{\prime }$.\newline In
Fig.(4-c), we fix $n=4$ and vary $\alpha $ as $0.0,0.1$ and $0.15$.
The present values of $w_{d}^{0},w_{d}^{\prime 0}$ are shown by
symbols on the lines. The circle symbol indicates the present value
($ w_{d}^{0}=-0.857,w_{d}^{\prime 0}=0.051$) in the case of $\alpha
=0.0$, the star shows ($w_{d}^{0}=-0.959,w_{d}^{\prime 0}=0.035$)
for $\alpha =0.1$ and the square represents
($w_{d}^{0}=-1.01,w_{d}^{\prime 0}=0.027$) for $\alpha =0.15$.
Increasing the interaction parameter $\alpha $ would lead to
decreasing the present value ($w^{0},w^{\prime 0}$) in $w-w^{\prime
}$ plane.

\section{Conclusion}

In this work, the interacting NADE model in non-flat universe has
been given. We studied the effect of spatial curvature $\Omega
_{k}$, interaction coefficient $\alpha $ and the main parameter of
NADE, $n$, On EoS parameter $w_{d}$ and deceleration parameter $q$.
We showed that in the early and present time for $\alpha _{1}=0.1,$
the phantom divide is not available for $n=3$ and it is achieved for
$n\geq 4$. By increasing $n$ and $\alpha $ in a closed universe, the
trend of $w_{d}$ decreases.   We obtained a minimum value for $n$ in
both early and present time, in order to that the NADE model crosses
the phantom divide. It was shown that the treatment of both
parameter $w_{d}$ and $q$ are dependent on the type of spatial
curvature. At last, we investigated the interacting NADE model in a
non-flat universe by means of statefinder diagnostic and
$w-w^{\prime }$ analysis viewpoints. Here we showed that the spatial
curvature can affect the evolution trajectories in ($s,r$) and ($
w_{d},w_{d}^{\prime }$) planes. Also the trajectories in these
planes can be affected by the model parameters of interacting NADE,
$n$ and $\alpha $.  In non flat universe, the statefinder trajectory
is discriminated by both $n$ and $\alpha$. It should be noted that
in the case of flat universe, only the parameter $\alpha$ can
discriminate the evolution trajectory in $s-r$ plane and $n$ can
only discriminate toady's value of $s$ and $r$(see Figs.(3 \& 4) of
\cite{zhang}). While in non flat universe, both parameteres $n$ and
$\alpha$ in addition to discrimination of \{$s, r$\} at present
time, can discriminate the evolution trajectory in $s-r$ plane (see
Figs.3-b \& 3-c). It is worthwhile to mention that all computations
is reduced to previous work in the limiting case of flat universe
\cite{zhang}.

\begin{center}
\begin{figure}[!htb]
\includegraphics[width=7cm]{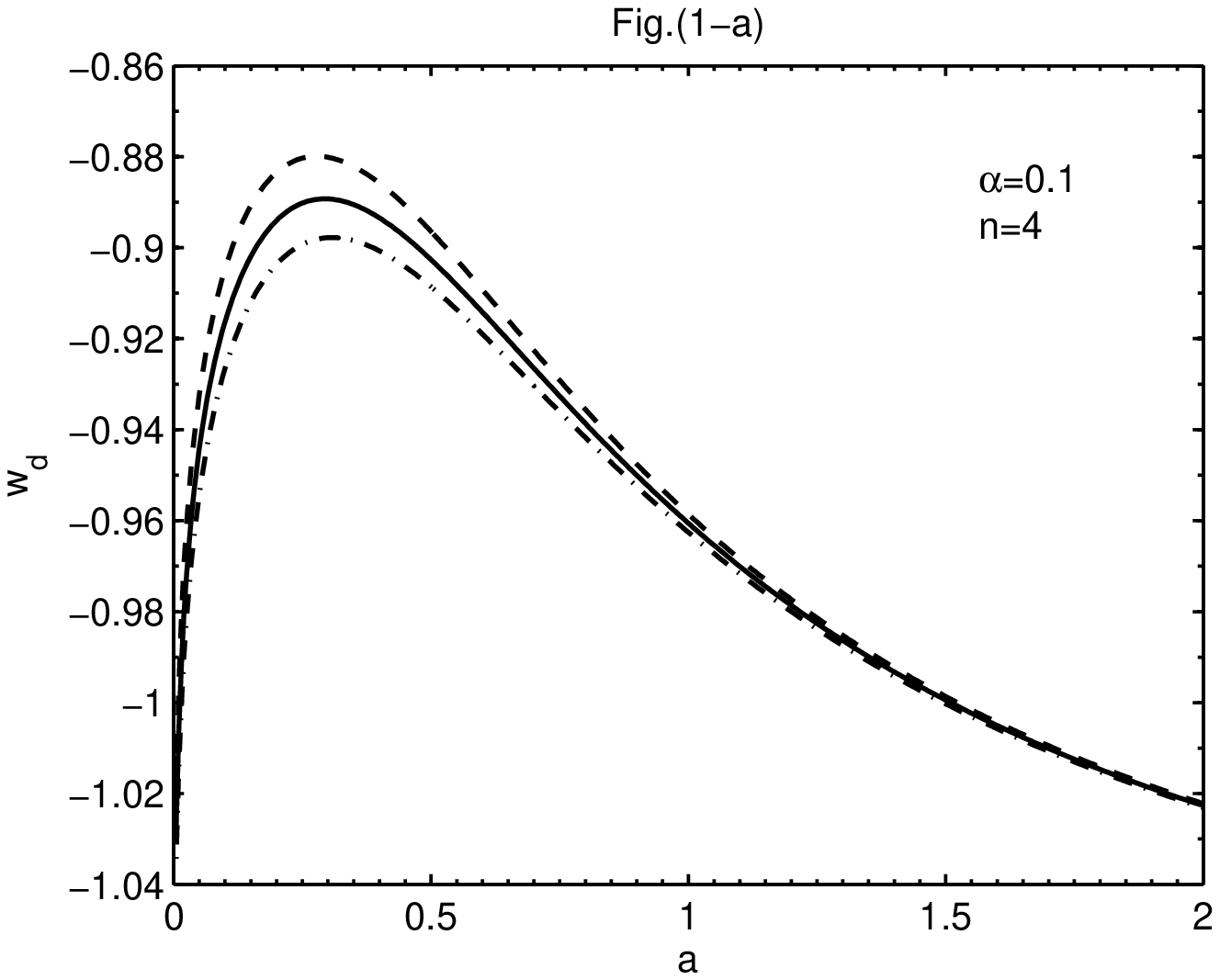}\newline
\vskip.3cm ~~~~~~\includegraphics[width=7cm]{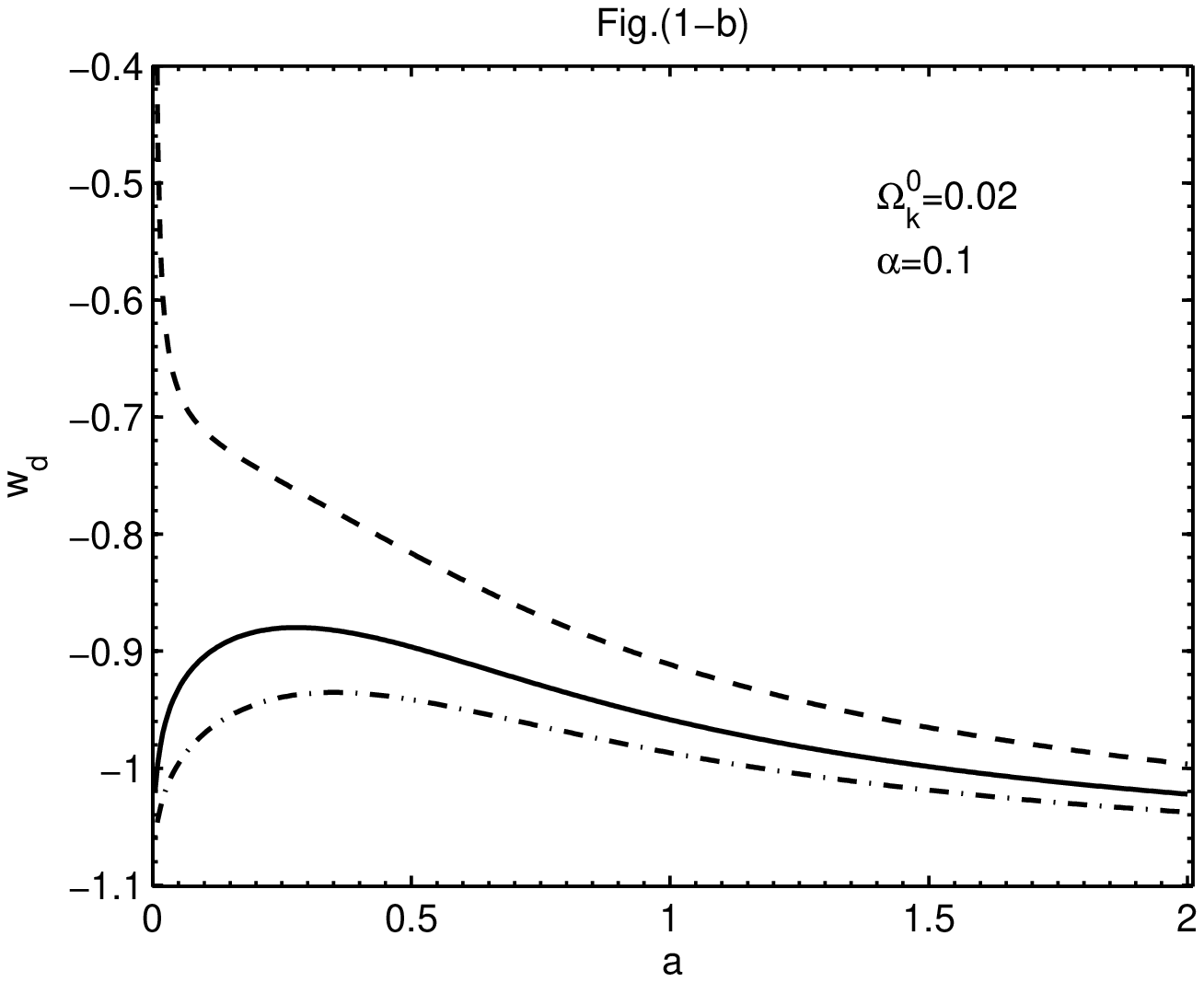} %
\includegraphics[width=7cm]{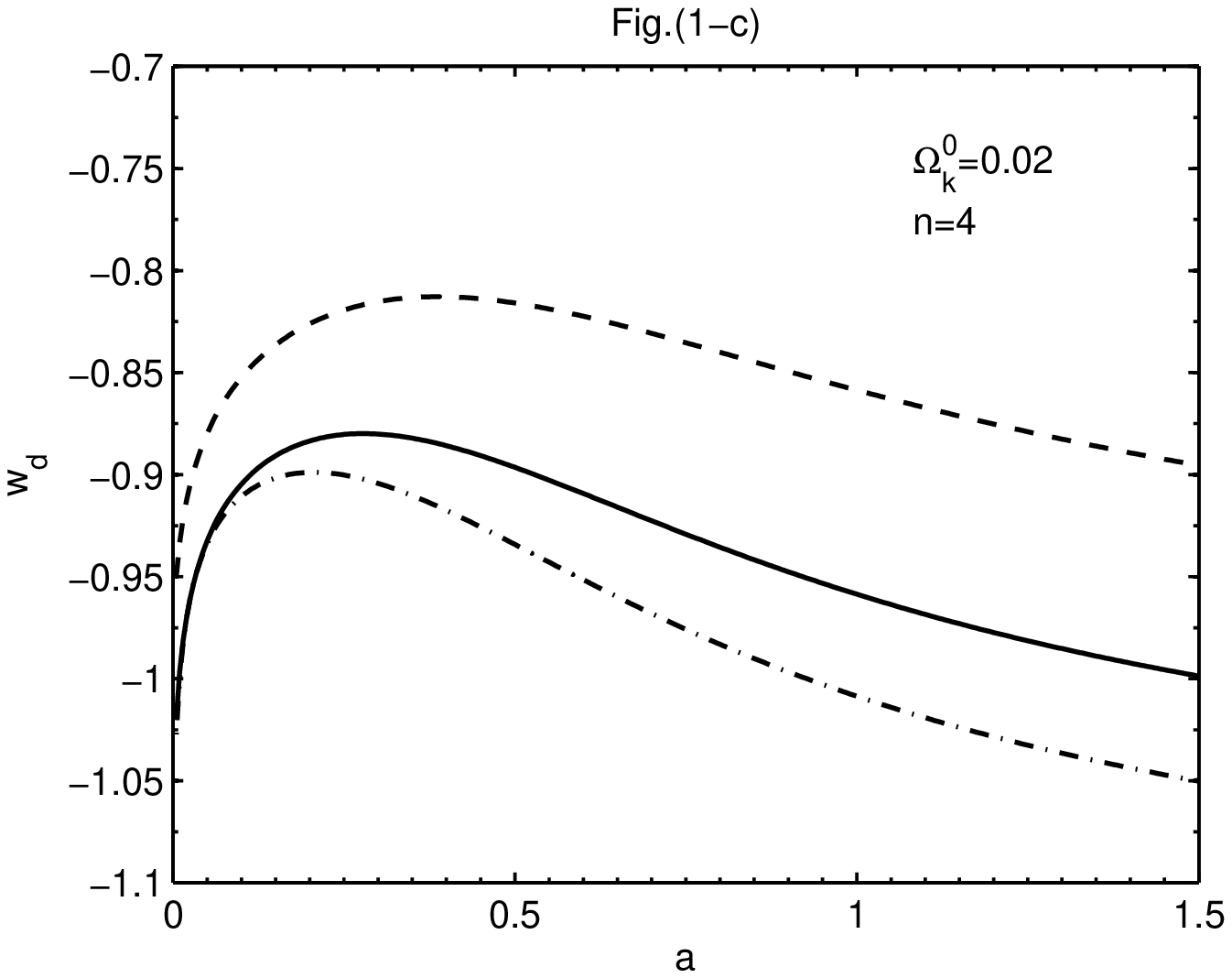} ~~~~~~
\caption{The evolution of EoS parameter, $w_d$, versus of $a$ for different
model parameters $n$, $\protect\alpha$ and different curvatures $\Omega_k$.
Fig.(1-a): Dependence of $w_d$ on $\Omega_k$ , for $n=4$ and $\protect\alpha%
=0.1$. The dashed, solid and dotted-dashed lines represent the closed, flat
and open universe, respectively. Fig.(1-b): the evolution of $w_d$ for $n=3$
(dashed line), $n=4$ (solid line) and $n=5$ (dotted-dashed line), by fixing $%
\protect\alpha=0.1$, in a closed universe . Fig.(1-c): the evolution of $w_d$
for $\protect\alpha=0.0$ (dashed line), $\protect\alpha=0.1$ (solid line)
and $\protect\alpha=0.15$ (dotted-dashed line) , by fixing $n=4$, in a
closed universe.\\[0pt]
}
\label{fig1}
\end{figure}
\end{center}

\newpage
\begin{figure}[!htb]
\includegraphics[width=7cm]{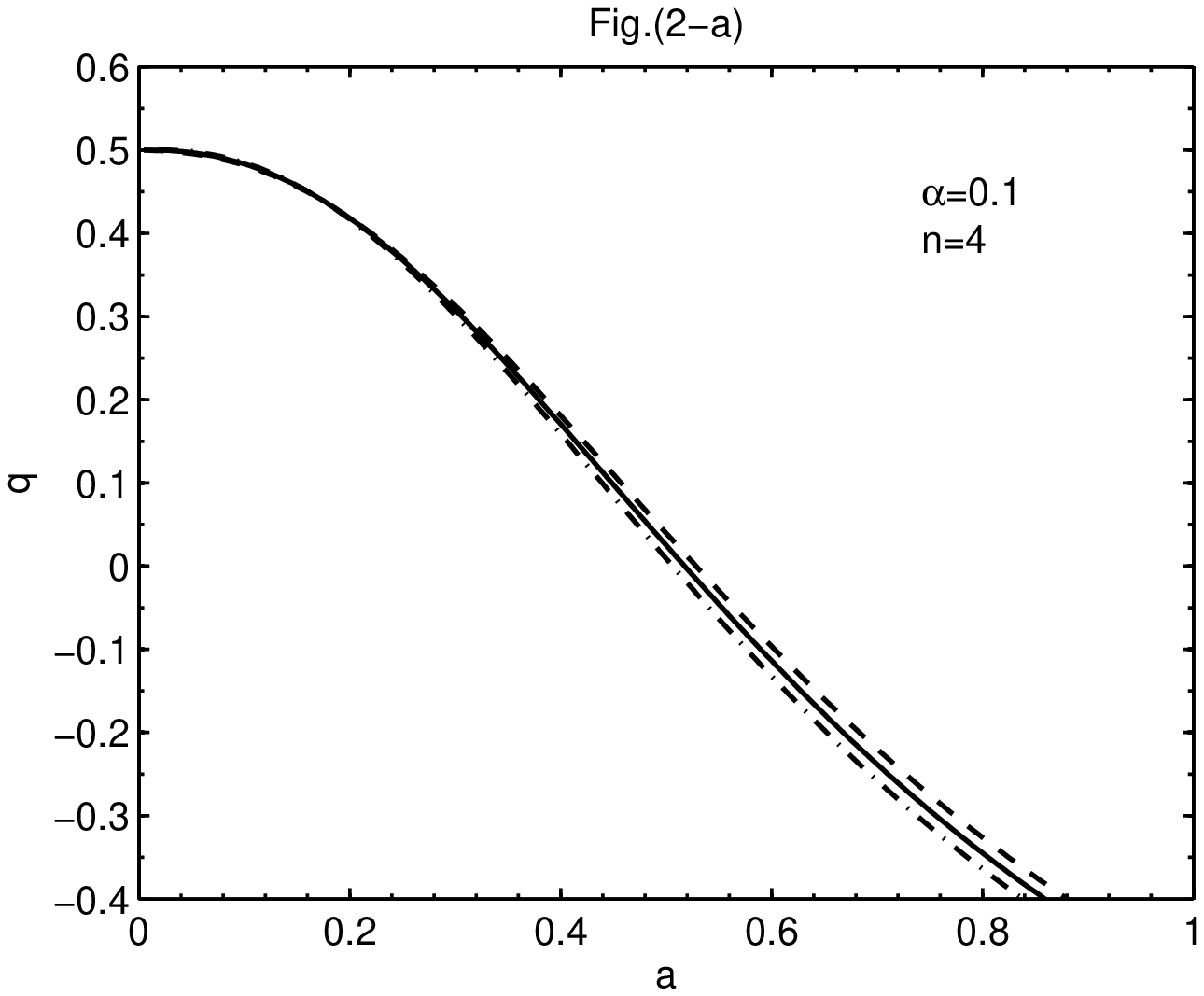}\newline
\vskip.3cm ~~~~~~\includegraphics[width=7cm]{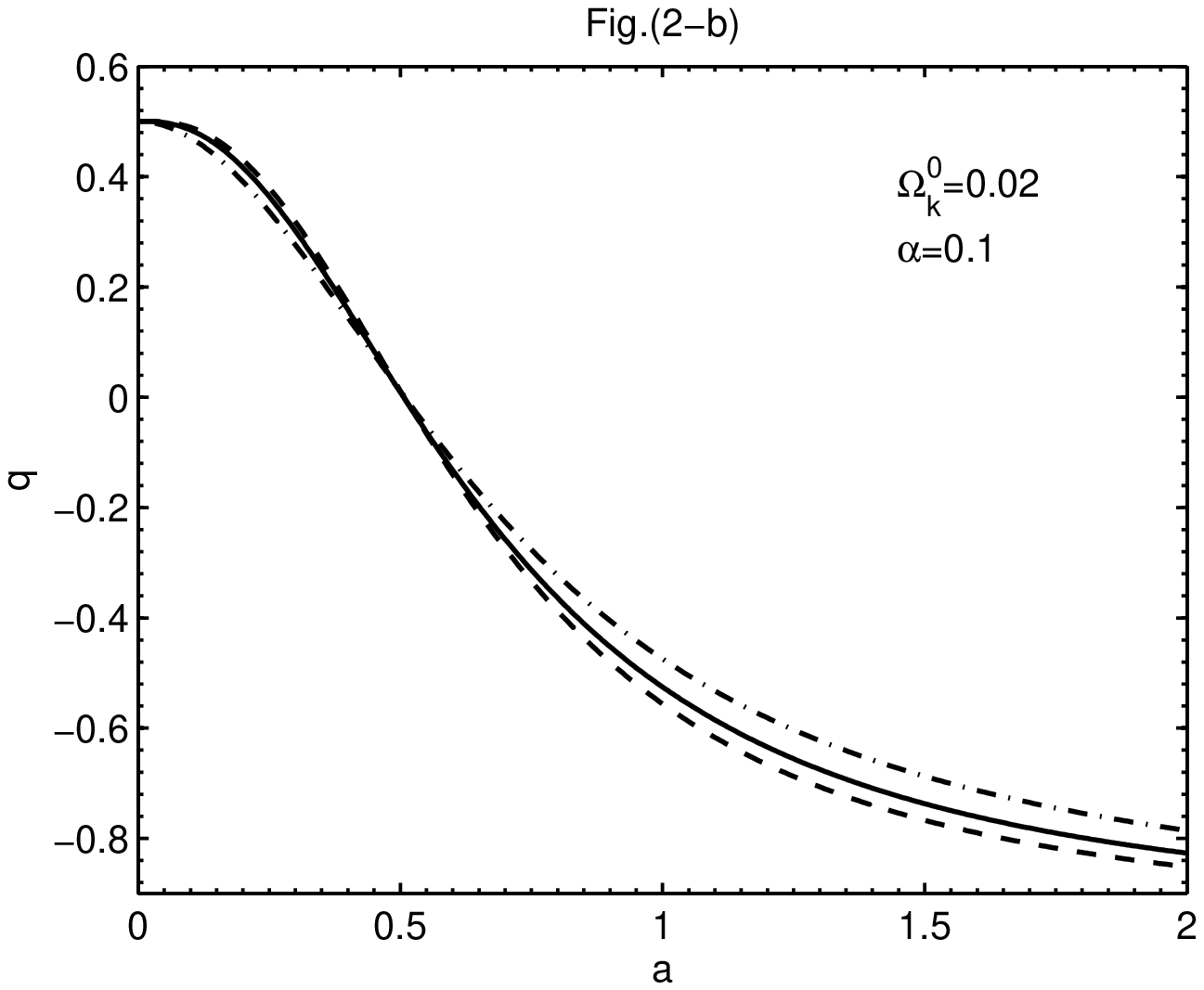} %
\includegraphics[width=7cm]{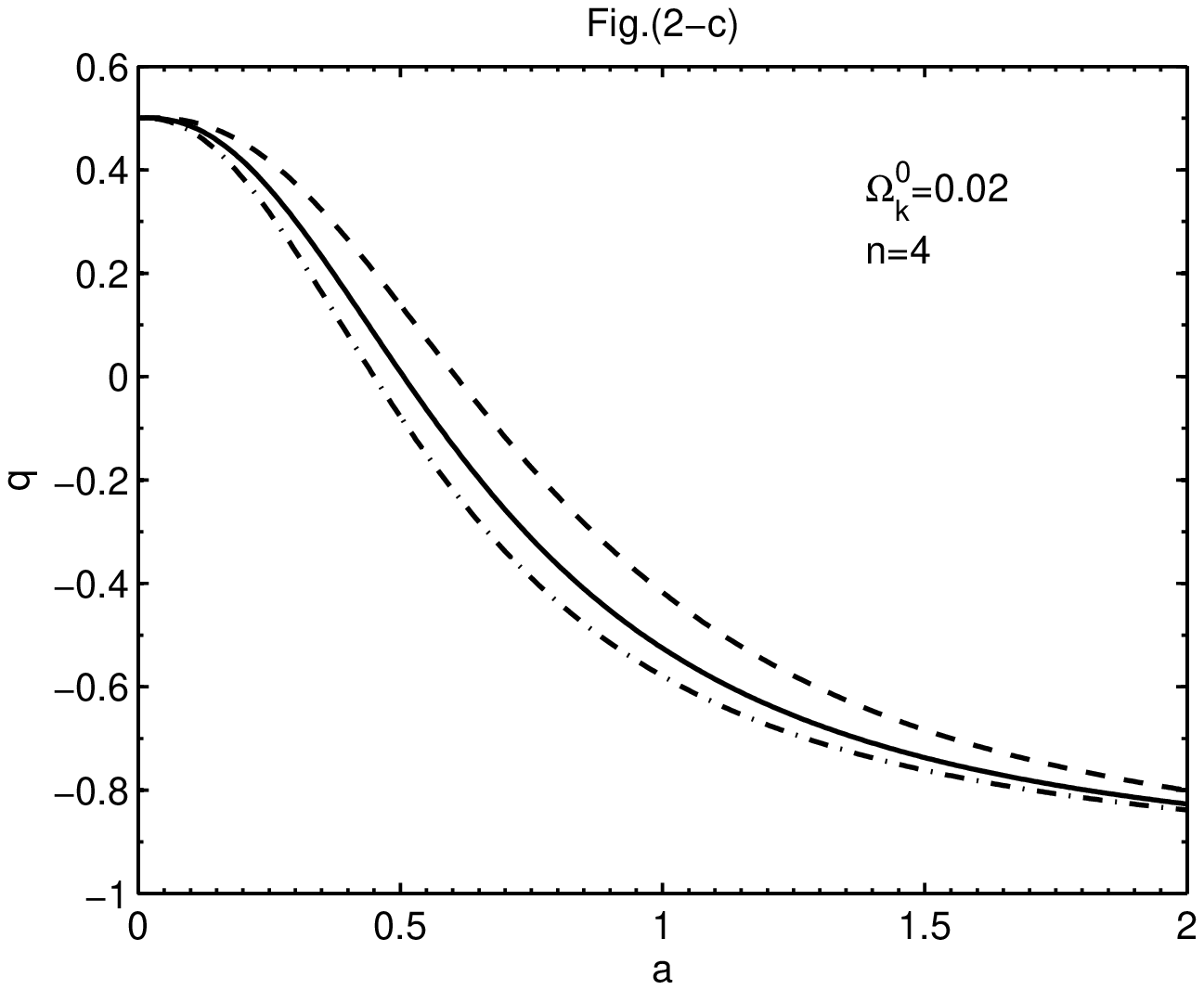}~~~ ~~~~~~
\caption{The evolution of $q$ versus of $a$ for different parameters $n$ and
$\protect\alpha$ and various contribution of spatial curvatures $\Omega_k$.
Fig.(2-a): Dependence of $q$ on $\Omega_k$ for values $n=4$ and $\protect%
\alpha=0.1$. The dashed, solid and dotted-dashed lines represent the open,
flat and closed universe, respectively. Fig.(2-b): the evolution of $q$
versus $a$, by fixing $\protect\alpha=0.1$, for $n=3$ (dotted-dashed line), $%
n=4$ (solid line) and $n=5$ (dashed line) in closed universe. Fig.(2-c): the
evolution of $q$ versus $a$, by fixing $n=4$, for $\protect\alpha=0.0$
(dotted-dashed line), $\protect\alpha=0.1$ (solid line) and $\protect\alpha%
=0.15$ (dashed line) in closed universe.\newline
}
\end{figure}
\newpage

\begin{center}
\begin{figure}[!htb]
\includegraphics[width=7cm]{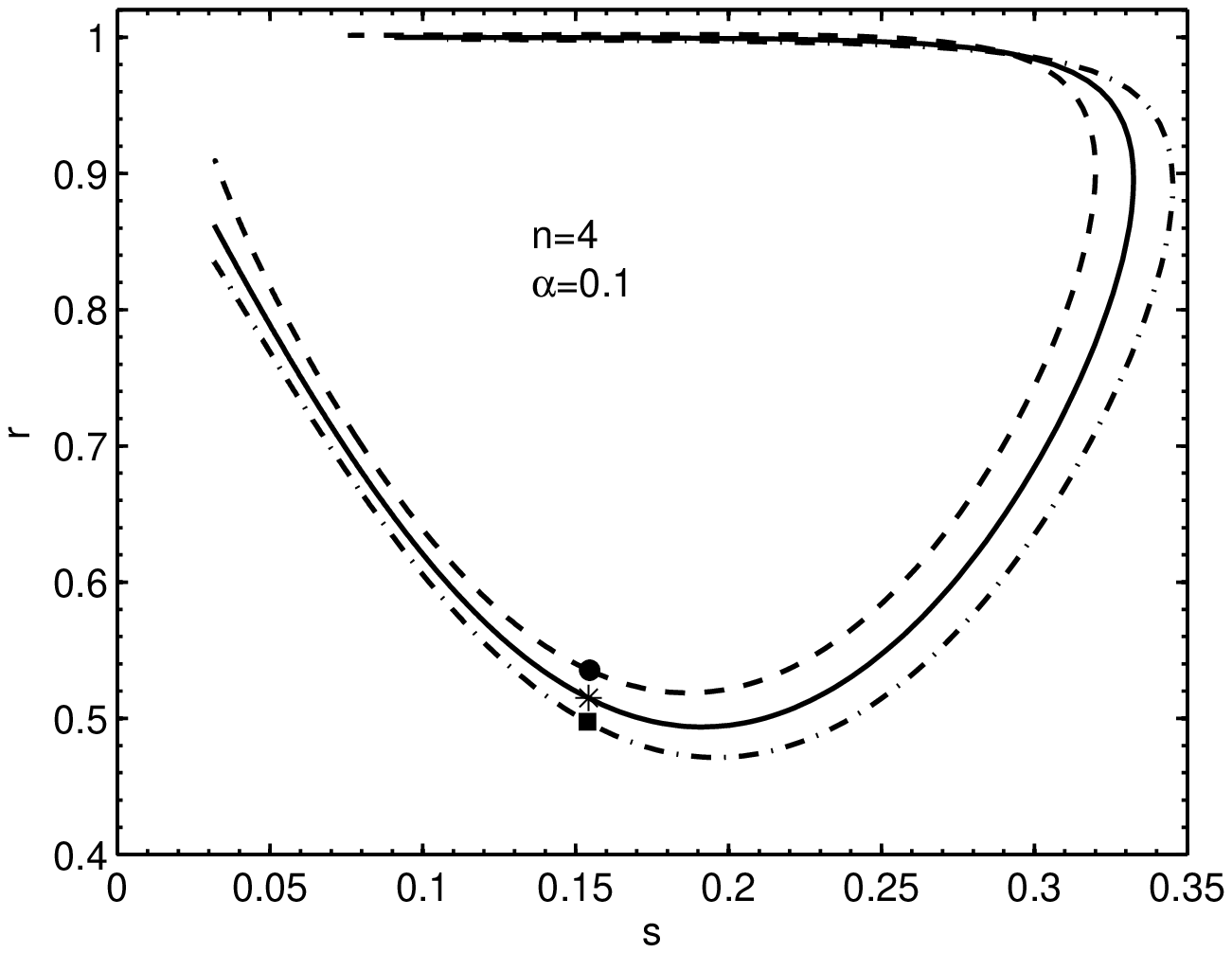}\newline
\vskip.3cm ~~~~~~\includegraphics[width=7cm]{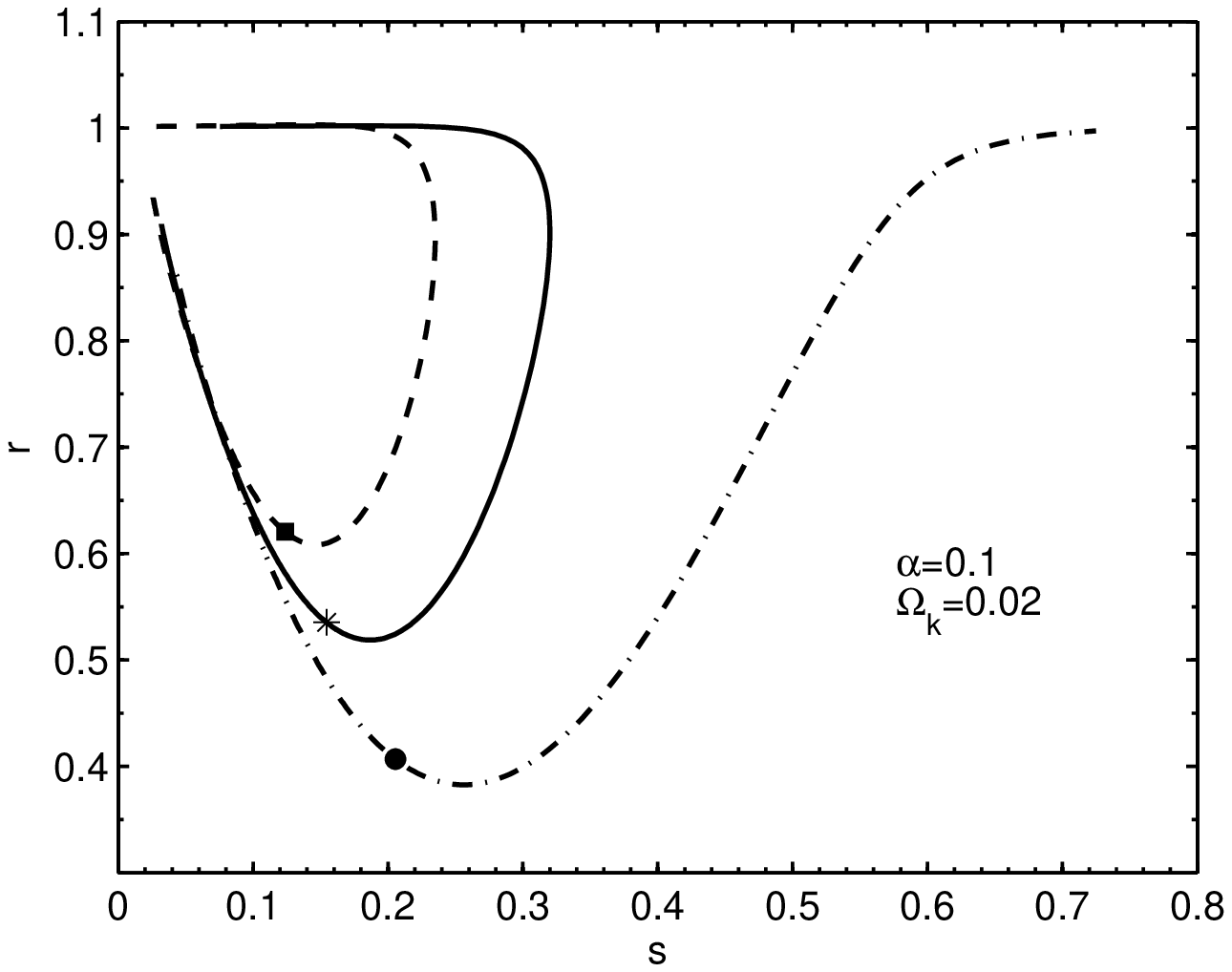} %
\includegraphics[width=7cm]{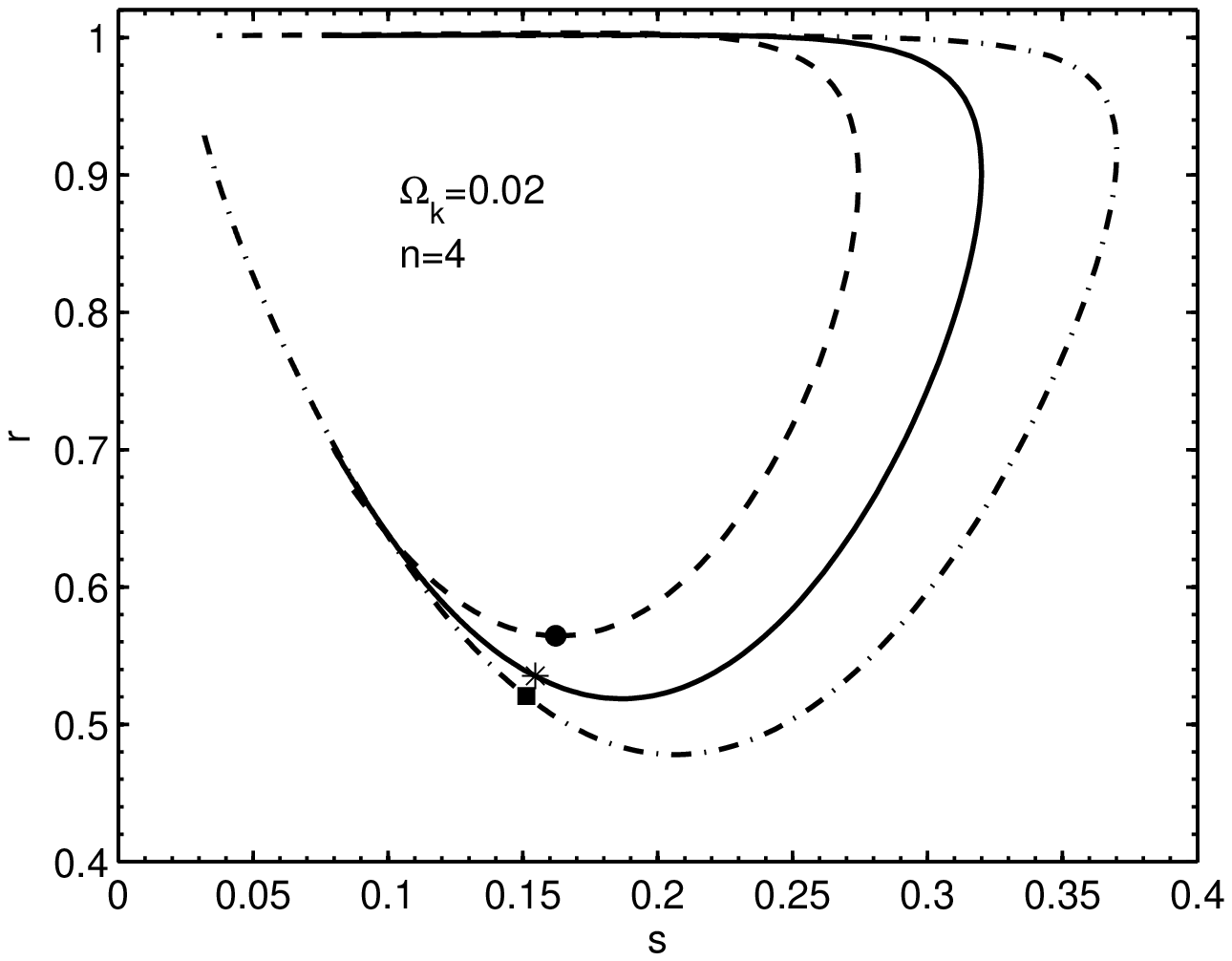} ~~~~~~
\caption{Fig.(3-a): Evolution trajectories of the statefinder in the $r-s$
plane for open (dotted-dashed line), flat (solid line) and closed universe
(dashed line). Square, circle and star symbols on the curves are the today's
values of the statefinder parameters ($s_0$,$r_0$). Here the model
parameters are $n=4$ and $\protect\alpha=0.1$.\\[0pt]
Fig.(3-b): Evolution trajectories of the statefinder in the $r-s$ plane for
different values of $n$ as 3 (dotted-dashed line), 4 (solid line) and 5
(dashed line) , by fixing $\protect\alpha=0.1$ in closed universe.
Fig.(3-c): Evolution trajectories of the statefinder in the $r-s$ plane for
different values of $\protect\alpha$ as $0.0$ (dotted-dashed line),$0.1$
(solid line) and $0.15$ (dashed line) , by fixing $n=4$ in closed universe. }
\label{fig3}
\end{figure}
\end{center}

\newpage

\begin{center}
\begin{figure}[!htb]
\includegraphics[width=7cm]{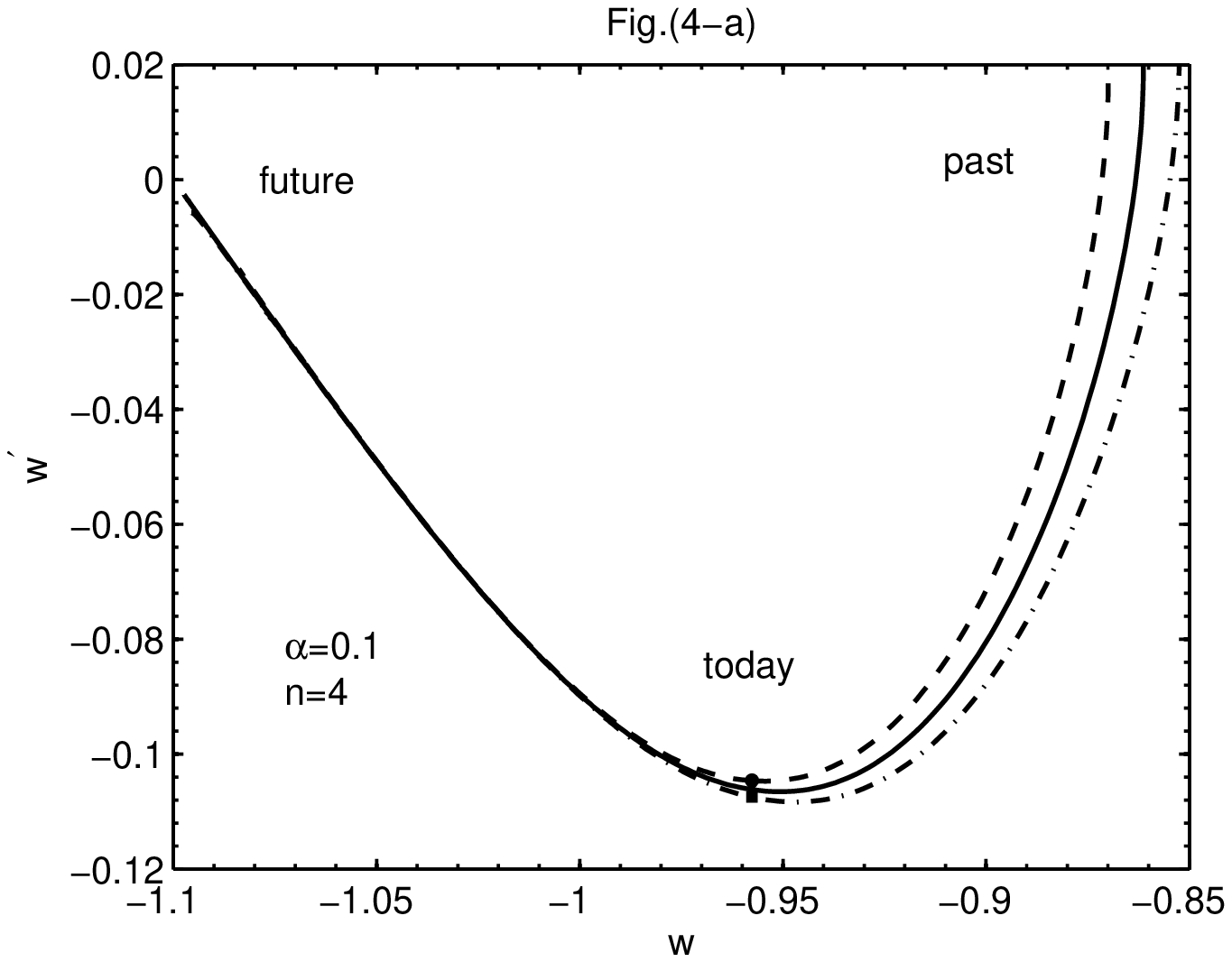}\newline
\vskip.3cm ~~~~~~\includegraphics[width=7cm]{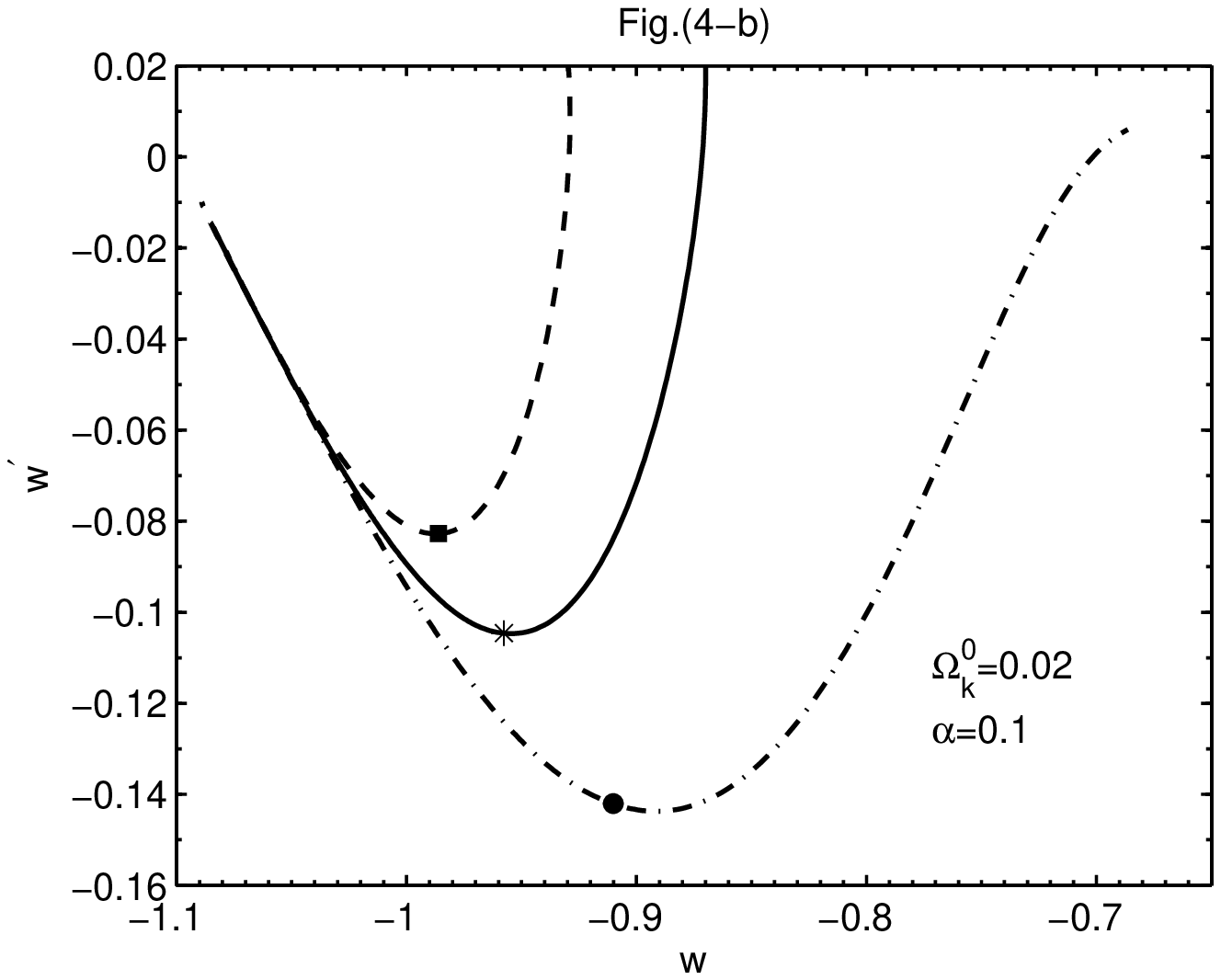} %
\includegraphics[width=7cm]{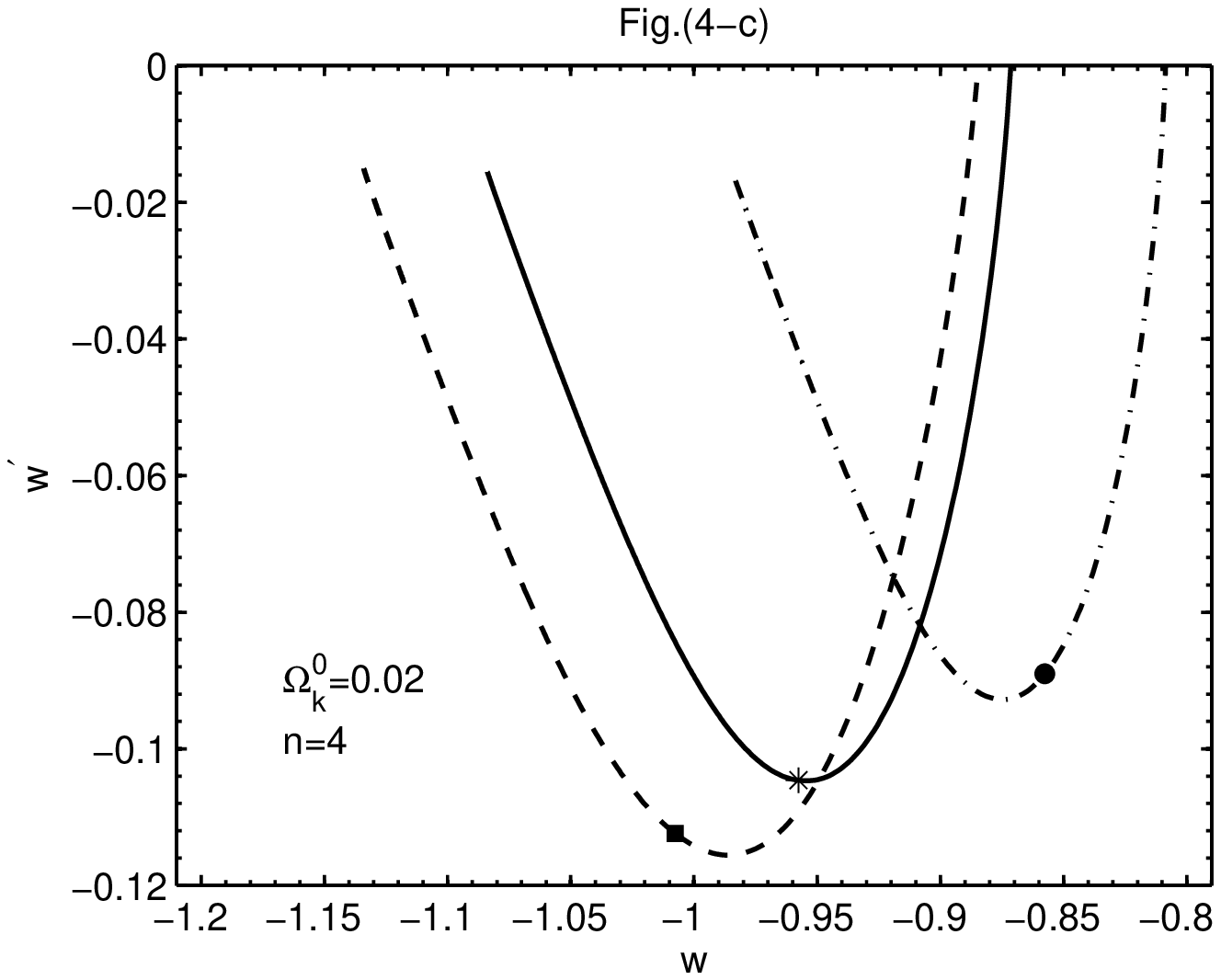} ~~~~~~
\caption{Fig.(4-a): the evolution trajectories in $w_d, w^{\prime}_d$ plane,
by fixing $n=4$ and $\protect\alpha=0.1$, for closed (dashed line), flat
(solid line) and open universe (dotted-dashed line). The present value of $%
w_d^0, w^{\prime 0}_d$ is indicated by symbols on the curves. Fig.(4-b):
Evolution trajectories in $w_d, w^{\prime}_d$ plane for different values of $%
n$ as 3 (dotted-dashed line), 4 (solid line) and 5 (dashed line) , by fixing
$\protect\alpha=0.1$ in closed universe. Fig.(4-c): Evolution trajectories
in $w_d, w^{\prime}_d$ plane for different values of $\protect\alpha$ as $%
0.0 $ (dotted-dashed line), $0.1$ (solid line) and $0.15$ (dashed line) , by
fixing $n=4$ in closed universe.}
\label{fig4}
\end{figure}
\end{center}

\end{document}